\documentclass{JHEP3}    
\usepackage{epsfig}
\usepackage{amsfonts}

\newcommand{\eref}[1]{(\ref{#1})}

\newcommand{\fref}[1]{Figure~\ref{#1}}
\newcommand{\cref}[1]{Chapter~\ref{#1}}
\newcommand{\beq}{\begin{equation}}
\newcommand{\eeq}{\end{equation}}
\newcommand{\ba}{\begin{array}}
\newcommand{\ea}{\end{array}}
\newcommand{\bcenter}{\begin{center}}
\newcommand{\ecenter}{\end{center}}

\def\IB{\relax\hbox{$\inbar\kern-.3em{\rm B}$}}
\def\IC{\relax\hbox{$\inbar\kern-.3em{\rm C}$}}
\def\ID{\relax\hbox{$\inbar\kern-.3em{\rm D}$}}
\def\IE{\relax\hbox{$\inbar\kern-.3em{\rm E}$}}
\def\IF{\relax\hbox{$\inbar\kern-.3em{\rm F}$}}
\def\IG{\relax\hbox{$\inbar\kern-.3em{\rm G}$}}
\def\IGa{\relax\hbox{${\rm I}\kern-.18em\Gamma$}}
\def\IH{\relax{\rm I\kern-.18em H}}
\def\IK{\relax{\rm I\kern-.18em K}}
\def\IL{\relax{\rm I\kern-.18em L}}
\def\IP{\relax{\rm I\kern-.18em P}}
\def\IR{\relax{\rm I\kern-.18em R}}
\def\IZ{\relax\ifmmode\mathchoice
{\hbox{\cmss Z\kern-.4em Z}}{\hbox{\cmss Z\kern-.4em Z}}
{\lower.9pt\hbox{\cmsss Z\kern-.4em Z}}
{\lower1.2pt\hbox{\cmsss Z\kern-.4em Z}}\else{\cmss Z\kern-.4em Z}\fi}
\def\II{\relax{\rm I\kern-.18em I}}

\def\sCC{{\kern 0.27em\vrule height1.45ex width0.03em depth0em
          \kern-0.30em\rm C}}
\def\C{{\mathchoice
  {\sCC}
  {\sCC}
  {\kern 0.225em \vrule height1.05ex width0.025em depth0em \kern-0.25em \rm C}
  {\kern 0.180em \vrule height0.78ex width0.02em depth0em \kern-0.2em \rm C}
        }}
\def\sHH{{\rm I\kern-.16em{}H}}
\def\H{{\mathchoice
  {\sHH}
  {\sHH}
  {\rm I\kern-.13em{}H}
  {\rm I\kern-.13em{}H} }}
\def\sNN{{\rm I\kern-.16em{}N}}
\def\N{{\mathchoice
  {\sNN}
  {\sNN}
  {\rm I\kern-.12em{}N}
  {\rm I\kern-.10em{}N} }}
\def\sPP{{\rm I\kern-.16em{}P}}
\def\P{{\mathchoice
  {\sPP}
  {\sPP}
  {\rm I\kern-.12em{}P}
  {\rm I\kern-.10em{}P} }}
\def\sQQ{{\kern 0.27em \vrule height1.45ex width0.03em depth0em
          \kern-0.30em \rm Q}}
\def\Q{{\mathchoice
        {\sQQ}
        {\sQQ}
  {\kern 0.225em \vrule height1.05ex width0.025em depth0em \kern-0.25em \rm Q}
  {\kern 0.180em \vrule height0.78ex width0.020em depth0em \kern-0.20em \rm Q}
        }}
\def\sRR{{\rm I\kern-0.16em{}R}}
\def\R{{\mathchoice
  {\sRR}
  {\sRR}
  {\rm I\kern-0.12em{}R}
  {\rm I\kern-0.10em{}R} }}
\def\sZZ{{\rm Z\kern-0.32em{}Z}}
\def\Z{{\mathchoice
  {\sZZ}
  {\sZZ} 
  {\rm Z\kern-0.3em{}Z}     %.3
  {\rm Z\kern-0.25em{}Z} }}  %.25
\def\ZZZ{{\rm Z\kern-0.24em{}Z}}
\def\sII{{\rm I\kern-0.16em{}I}}
\def\I{{\mathchoice
  {\sII}
  {\sII}
  {\rm I\kern-0.12em{}I}
  {\rm I\kern-0.10em{}I} }}

\def\inbar{\,\vrule height1.5ex width.4pt depth0pt}
\font\cmss=cmss10 \font\cmsss=cmss10 at 7pt

\def\smiley{\hbox{\large$\bigcirc$\hspace{-0.80em}\raise.2ex
\hbox{$\cdot\cdot$}\kern-.61em\lower.2ex\hbox{\scriptsize$\smile$}}\ }
\def\frowny{\hbox{\large$\bigcirc$\hspace{-0.80em}\raise.2ex
\hbox{$\cdot\cdot$}\kern-.635em\lower.2ex\hbox{\scriptsize$\frown$}}\ }

\def\I{{\rlap{1} \hskip 1.6pt \hbox{1}}}

\makeatletter
\let\hangafter\@hangfrom
\makeatother

\newcommand{\be}{\begin{equation}}
\newcommand{\ee}{\end{equation}}
\newcommand{\bea}{\begin{eqnarray}}
\newcommand{\eea}{\end{eqnarray}}
\newcommand{\bean}{\begin{eqnarray*}}
\newcommand{\eean}{\end{eqnarray*}}

\preprint{MIT-CTP-3508}

\title{On the fate of tachyonic quivers}

\author{Sebasti\'{a}n Franco and Amihay Hanany
\footnote{
Research supported in part by the CTP and the LNS
of MIT and the U.S. Department of Energy under cooperative agreement
$\#$DE-FC02-94ER40818, and by BSF -- American-Israeli Bi-National Science 
Foundation. A. H. is also supported by the Reed Fund Award and 
a DOE OJI award.}
\\
~\\
Center for Theoretical Physics,
\\ Massachusetts Institute of Technology,\\
Cambridge, MA 02139, USA.\\
\email{sfranco, hanany@mit.edu}
}

\abstract{We study gauge theories on the world-volume of D3-branes probing singularities. Seiberg duality can be realized 
as a sequence of Picard-Lefschetz monodromies on 3-cycles in the mirror manifold. In previous work, the precise meaning of 
gauge theories obtained by monodromies that do not correspond to Seiberg duality was unclear. Recently, it was pointed out 
that these theories contain tachyons, suggesting that the collection of branes at the singularity is 
unstable. We address this problem using $(p,q)$ web techniques. It is shown that theories with tachyons appear whenever 
the $(p,q)$ web contains crossing legs. A recent study of these theories with tachyons using exceptional collections 
proposed the notion of ``well split condition.'' We show the equivalence between the well split condition and the absence 
of crossing legs in the $(p,q)$ web. The $(p,q)$ web has a natural resolution of crossing legs which was first studied in 
the construction of five dimensional fixed points using branes. Exploiting this result, we propose a generic procedure which 
determines the quiver that corresponds to the stable bound state of D-branes that 
live on the singularity after the monodromy. This set is generically larger than the original set, meaning 
that there are extra massless gauge fields and matter fields in the quiver. Alternatively, one can 
argue that since these gauge and matter fields are initially assumed to be absent, the theory exhibits tachyonic 
excitations. We illustrate our ideas in an explicit example for D3-branes on a complex cone over $dP_1$, 
computing both the quiver and the superpotential.}
\begin{document}

%%%%%%%%%%%%%%%%%%%%%%%%%%%%%%%%%%%%%%%%%%%%%%%%%%%%%%

%=====================================================
\section{Introduction}
%=====================================================

There are several ways to engineer supersymmetric gauge theories using the low energy limit of String Theory.
One of them is by probing singularities with a stack of D-branes. When a non-compact Calabi-Yau
is probed with D3-branes, the result is an $\mathcal{N}=1$, 3+1 dimensional gauge theory. 
More precisely, for every given singularity there is an infinite number of field theories
that are related in the infrared by generalized Seiberg dualities \cite{Seiberg:1994pq}. 

When field theories are constructed in this way, it is possible to give a geometric 
interpretation to duality transformations. In particular, Seiberg duality can be
obtained from Picard-Lefschetz (PL) monodromies of 3-cycles in the mirror manifold 
\cite{Ooguri:1997ih,Feng:2001bn,Cachazo:2001sg}.

An interesting family of singularities is the one of complex cones over del Pezzo surfaces. The gauge theories
associated to these geometries have been shown to exhibit a very rich physics, including chaotic RG flows
and duality walls \cite{Hanany:2003xh,Franco:2003ja,Franco:2003ea,Franco:2004jz}.

The computation of quiver theories for D-branes on singularities was substantially simplified in 
\cite{Hanany:2001py} using the $(p,q)$ web techniques of \cite{Aharony:1997ju, Aharony:1997bh}. 
The methods in \cite{Hanany:2001py} showed how one can compute the quiver matrix using data which can be read from the $(p,q)$ web which describes the geometry under study.
Working in this context, it was realized in \cite{Feng:2001bn} that Seiberg duality transformations form a subgroup of the 
larger set of Picard-Lefschetz monodromies on $(p,q)$ 7-branes. In particular it was demonstrated that there are PL 
monodromies which are not Seiberg duality. Furthermore, it was shown explicitly in this language how a single action of Seiberg duality is composed of few monodromies \cite{Feng:2002kk}. For this reason, theories that result from non-Seiberg monodromies were called {\bf fractional Seiberg duals}. Few questions remain open: How are these fractional Seiberg duals related to the usual Seiberg duals? Do 
they fall into the same universality class in the infrared as for the case of ordinary Seiberg duals (i.e. in what sense, if any, 
are they dual)? Do two theories which are related by fractional Seiberg duality have the same moduli space of vacua? The 
same spectrum of gauge invariant operators? etc.

A prescription for deriving the quiver that results from fractional Seiberg duality is at hand. It uses intersections of 
3-cycles in the mirror manifold. However, it is not clear how to compute the superpotential for such a quiver. This is 
another puzzle. Parallel discussions, in terms of exceptional collections, appeared subsequently in a series of papers by 
Herzog \cite{Herzog:2003zc,Herzog:2004qw} \footnote{A different perspective in the study of Seiberg duality was pursued in 
\cite{Berenstein:2002fi,Braun:2002sb,Mukhopadhyay:2003ky}, where it was realized as a tilting equivalence of the quiver 
derived category.}. The main focus of the work that followed \cite{Feng:2002kk} was on the 
identification of theories that are not Seiberg duals and on looking for a criterion that enables us to consistently 
restrict our attention to Seiberg dual theories, if it exists at all. It was argued in various examples that quivers for 
fractional Seiberg duals sometimes lead to IR theories with gauge invariant chiral operators of negative R-charge. That is, if one follows the same techniques as 
for usual Seiberg duals, such negative R-charges appear. One is lead to conclude that even though fractional Seiberg duals 
look like ordinary quiver gauge theories in the UV, they are in fact inconsistent or possibly incomplete.

Recently, an interesting observation about del Pezzo quivers was made by Aspinwall and Melnikov \cite{Aspinwall:2004vm}. 
The authors emphasized a distinguishing characteristic of some of the quivers for the del Pezzo manifolds that turn out to 
be the fractional duals mentioned above. Some of the bifundamental fields in these models are induced by $Ext^3$ groups 
and are thus tachyonic \footnote{A related discussion of these states can be found in \cite{Herzog:2004qw}.}. The authors proved 
that a quiver is free of bifundamental tachyons if and only if 
$Ext^p(L_i,L_j)=0$ for $i \neq j$ with $p \geq 3$. They also showed that one can get rid of the undesired tachyons 
by performing appropriate mutations (which is equivalent to the statement that these models are obtained from tachyon-free 
quivers by performing the inverse mutations). This certainly adds to the understanding of the gauge theories obtained from 
D-branes on complex cones over del Pezzo surfaces but raises natural questions, closely related to the ones we posed 
above. How should we interpret (if such an interpretation actually exists) the tachyonic quivers? Do they decay by the 
condensation of the tachyons or should we look for a different, stable configuration of D-branes at the singularity?

One can ask in what sense does a tachyon appear in such quiver theories? Does one of the bifundamental fields have a 
negative mass squared? This is hard to do since all of the bifundamental fields in a quiver theory that results from an 
exceptional collection are chiral. Therefore mass terms cannot be written for such fields. What is the gauge theory 
indication of an instability? 

In this paper we will use the $(p,q)$ web techniques of \cite{Hanany:2001py} to provide a new perspective into the problem 
of these tachyonic quivers or fractional Seiberg duals. As we will see, the webs associated to fractional dual theories are
characterized by having {\bf crossing external legs}. This seems to be one of the simplest ways of identifying these models. We will exploit the information in the webs further, and use them to give a concrete proposal for the stable quiver which results after the monodromy is performed. We will see that in these models, given a set of asymptotic $(p,q)$ 7-branes, the minimal resolution of the corresponding singularity is of greater genus (here genus refers to the number of compact faces in the $(p,q)$ web) than before applying the monodromy. That is, the appropriate stable set of branes corresponds to a quiver with more gauge groups and more matter fields than in the original one. This argument is general enough such that it can be applied to quiver theories 
for D3-branes on geometries with an arbitrary number of collapsing 4-cycles (as above, this number is simply given by the genus of the corresponding web) and not only to del Pezzos which have only one collapsing 4-cycle. 

The organization of the paper is as follows. In Section \ref{section_webs}, we review the $(p,q)$ web description
of singularities. In Section \ref{section_crossing_legs}, we discuss the phenomenon 
of crossing legs in $(p,q)$ webs and state our proposal for the stable quiver after monodromies. 
In Section \ref{section_well_split}, we prove that the well split condition derived from exceptional collections 
to restrict the theories under study is actually equivalent to the absence of crossing legs in the associated web.
Section \ref{section_old_puzzles} shows how some seemingly unrelated difficulties in interpreting some quivers are in 
fact different manifestations of the same problem. Finally, guided by the perspective of crossing of external legs in
$(p,q)$ webs, we propose in Section \ref{section_higher_genus} that the stable set of D-branes for a fractional dual 
corresponds to a higher genus $(p,q)$ web. We construct this higher genus quiver explicitly for a model obtained after performing 
a Picard-Lefschetz monodromy on a $dP_1$ quiver. We derive this theory using $(p,q)$ webs 
machinery. There are no other
straightforward methods to compute quivers for non-orbifold singularities with more than 
one collapsing 4-cycle. 
This gauge theory for a genus 2 $(p,q)$ web is the main result of our paper. Finally, we discuss in Section \ref{section_negative_R} the existence of negative
R-charges in some fractional duals.

%=====================================================
\section{$(p,q)$ web description of singularities}
%=====================================================

\label{section_webs}

In \cite{Aharony:1997ju, Aharony:1997bh}, $(p,q)$ webs were introduced for studying five dimensional
gauge theories with 8 supercharges. They are configurations of 5-branes in Type IIB string theory. The branes
share 4+1 dimensions in which the field theory lives, and the physics is determined by the non-trivial configuration
in a transverse plane. 

There is an alternative interpretation of $(p,q)$ webs as toric skeletons describing toric varieties. The precise correspondence was worked out
in \cite{Leung:1997tw}. In this context, there is a $T^2$ fibration over every point of a web, with an $S^1$ going to zero size at each segment and the 
entire $T^2$ degenerating at the nodes. From this point of view, $(p,q)$ webs are reciprocal to ordinary toric diagrams. We will use this 
correspondence in Section \ref{section_higher_genus}.

Blowing-up a point (i.e. replacing it by a 2-sphere) corresponds in this language to replacing
it by a segment, whose length is given by the size of the sphere. Different $(p,q)$ webs can describe the same geometry. For example,
for toric del Pezzo surfaces $dP_n$ ($n \leq 3$), the $SL(3,\IC)$ symmetry of $\mathbb{P}^2$ can be used to place the 
blown-up points at any desired position.

\begin{figure}[ht]
  \epsfxsize = 4.5cm
  \centerline{\epsfbox{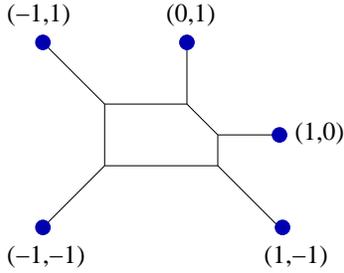}}
  \caption{A typical $(p,q)$ web. From a five dimensional point of view, it describes a supersymmetric gauge theory with 8 supercharges, an $SU(2)$ gauge group, and 1 flavor. If interpreted as a toric skeleton, it corresponds to $dP_2$. The  $(p,q)$ charges of the external legs are shown. Blue circles indicate the $(p,q)$ 7-branes at the ends of external legs.}
  \label{web_example}
\end{figure}

The use of webs to describe geometries can be extended by attaching external legs to $(p,q)$ 7-branes \cite{DeWolfe:1999hj}.
In this case, the legs no longer extend to infinity, permitting the study of cases that would otherwise have crossing legs. In addition, 
7-branes make the global symmetries of the corresponding five dimensional field theories manifest. We present 
a typical $(p,q)$ web example in \fref{web_example}.

In what follows, we will be interested in gauge theories on the world-volume of D3-branes probing singularities. The singularities we will focus on 
are complex cones over 2-complex dimensional toric compact manifolds $Y$ \footnote{In fact, although $(p,q)$ webs have a
natural interpretation in terms of toric geometry, their application extends to the determination of quivers
for non-toric singularities, as shown in \cite{Hanany:2001py}.}. These compact manifolds admit a $(p,q)$ web description.
The $(p,q)$ webs encode the information about the degenerate fibers of the mirror manifold. This fact was exploited in \cite{Hanany:2001py},
where a procedure for extracting the quiver of gauge theories in the world-volume of the D3-brane probes was developed. The number of gauge groups of the
quiver theory is given by the Euler characteristic of $Y$, which in terms of the Betti numbers is given by

\beq
\chi(Y)=b_0+b_2+b_4
\eeq

$\chi(Y)$ can be immediately read from the $(p,q)$ web
and is given by the number of nodes. In addition, we will refer to the number of compact faces of a $(p,q)$ web as its 
genus, which is equal to the number of collapsing 4-cycles.

For genus 1 webs, there is a one to one correspondence between web nodes (gauge groups) and external legs. The number of bifundamental fields between 
a pair of nodes $i$ and $j$ is given by the intersection of the corresponding 3-cycles in the mirror manifold, which in this case becomes

\beq
C_i \cdot C_j = \det \left( \begin{array}{cc} p_i & q_i \\ p_j & q_j \end{array}\right)
\label{intersection}
\eeq

We refer the reader to \cite{Hanany:2001py,Franco:2002ae,Feng:2004uq}, for a more detailed explanation of the connection between $(p,q)$ webs 
and toric geometry, and the computation of quivers on D-brane probes.

Picard-Lefschetz monodromy on the 3-cycles of the mirror geometry has a straightforward implementation as an action
on the $(p,q)$ charges of the 7-branes. The charges of a 7-brane that is moved around another one with charges $(p_i,q_i)$
are modified by multiplying them by the monodromy matrix

\beq
M(p_i,q_i)=\left(\begin{array}{rcr} 1+p_iq_i  & \ & -p_i^2 \\ q_i^2 & & 1-p_iq_i \end{array}\right)
\label{monodromy}
\eeq

Operating on $(p,q)$ 7-branes with \eref{monodromy} and its inverse is equivalent to acting with left and right mutations on 
the corresponding exceptional collections.

%=====================================================
\section{Brane crossing}
%=====================================================

\label{section_crossing_legs}

\begin{figure}[ht]
  \epsfxsize = 5.5cm
  \centerline{\epsfbox{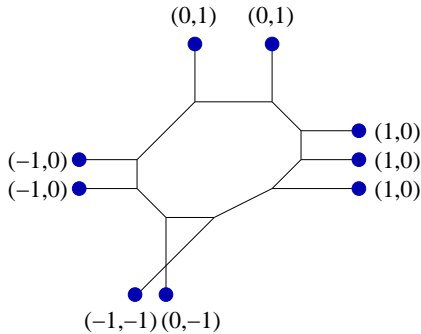}}
  \caption{Brane crossing in an attempt to construct a $(p,q)$ web for a five dimensional gauge theory with 8 supercharges, a gauge group $SU(2)$, and 5 flavors in the fundamental representation. The genus of the curve is the number of compact faces in the web. Here it corresponds (before considering the addition of a face due to
the crossing legs) to the rank of the $SU(2)$ gauge group, namely 1.}
  \label{web_5d}
\end{figure}

The application of $(p,q)$ web techniques to the determination of quiver gauge theories on D3-branes on singularities \cite{Hanany:2001py} identifies 
Seiberg duality as a Picard-Lefschetz monodromy \cite{Feng:2001bn}. The action of some Picard-Lefshetz monodromies leads to the phenomenon of {\bf brane 
crossing}. Brane crossing was first observed in \cite{Aharony:1997ju} when trying to construct a brane configuration for a five dimensional gauge theory with 8 supercharges that has more flavors than some critical value. \fref{web_5d} shows an attempt to construct a brane configuration for an $SU(2)$ gauge theory with 5 flavors. An immediate consequence is the appearance of crossing branes. More generally, for a brane construction of an $SU(n)$ gauge theory, if the number of flavors is more than $2n$ then the phenomenon of crossing branes is unavoidable. 

\begin{figure}[ht]
  \epsfxsize = 5.5cm
  \centerline{\epsfbox{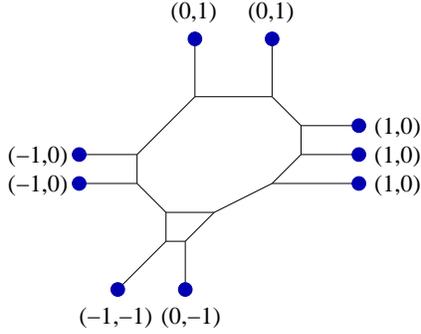}}
  \caption{Web diagram originated from the configuration in \fref{web_5d} after resolving the intersection point. The genus of the new $(p,q)$ web is 2.}
  \label{web_5d_crossing}
\end{figure}

Web diagrams indicate a possible solution to brane crossing. In fact, one can resolve the intersection point by 
replacing it by a segment, respecting the rules for the construction of a $(p,q)$ web. After doing so, we obtain a 
$(p,q)$ web whose genus has been increased. For the example of \fref{web_5d}, a resolution of the $(p,q)$ web 
diagram leads to the brane configuration of $SU(3)$ gauge theory with 5 flavors as can be seen in \fref{web_5d_crossing}.

From a five dimensional point of view, we start with a brane configuration describing an $SU(2)$ gauge theory, namely 
with rank 1, and at some high vev for the scalar field we discover another brane which corresponds to an increase of the 
rank to 2, namely an $SU(3)$ gauge group. Viewing this from the $SU(3)$ gauge theory point of view, there are three 
adjoint vevs, $\phi_1,\phi_2,\phi_3$ which satisfy the condition $\sum_i\phi_i=0$. The two relevant scales are the 
differences $\phi_1-\phi_2$ and $\phi_2-\phi_3$. When one of these scales is much higher than the other scale, the 
effective theory can be described by an $SU(2)$ gauge theory with a vev given by the smaller scale. This is the theory 
that we think we have at energies smaller than the high scale. However, once we start probing at energies comparable to 
the high scale, the $SU(2)$ description is not valid anymore and we need to refer to the more complete $SU(3)$ theory with 
5 flavors. The $SU(2)$ theory presents some instability which arises due to the fact that the description is lacking the 
additional states coming from the $SU(3)$ theory.

This is the essence of our proposal. In what follows we will apply this principle to resolve quivers with an 
inconsistency -- fractional Seiberg duals, ill split quivers, or tachyonic quivers -- into consistent quivers by resolving 
their corresponding web diagram along the lines described in the previous paragraphs.

Given a quiver that corresponds to a $(p,q)$ web with no brane crossing, we can perform a PL transformation. We have two 
cases to consider: If the resulting $(p,q)$ web has no crossing legs then the associated quiver will correspond to a 
Seiberg dual on the node over which the monodromy was performed \footnote{We want to remind the reader that, in some cases, 
Seiberg duality corresponds to a sequence of more than one PL monodromy \cite{Feng:2002kk}.}. For such a case, we start with a consistent quiver 
and end up again with a consistent quiver.
If the resulting $(p,q)$ web has brane crossing like that of \fref{web_5d}, then we need to resolve it and arrive at a 
$(p,q)$ web with higher genus. The resolution is dictated by the rules of construction of $(p,q)$ webs. 
Therefore we get a unique prescription for the resulting toric diagram and hence the resulting singular geometry. The new 
$(p,q)$ web corresponds to a different toric diagram, a different singular manifold, and will give rise to a different 
quiver gauge theory. For this quiver theory, higher genus means more gauge groups and more matter fields.

In Section \ref{section_higher_genus}, we will interpret this proposal 
from the point of view of the quiver gauge theory on D3-brane probes.

%===================================================================================
\section{The well split condition from $(p,q)$ webs: tachyons from crossing legs}
%===================================================================================

\label{section_well_split}

In \cite{Feng:2002kk}, it was realized that certain condition on a set of 7-branes guarantees that 
Seiberg duality in any node can be obtained by a sequence of PL monodromies. Exactly this same 
condition was identified in \cite{Herzog:2003zc} in the context of exceptional collections, and was dubbed
{\bf well split} condition. Quivers that are not well split are labeled ill split. A well split quiver is 
such that for any node $i$, all the in-going nodes into $i$ are placed to the left of $i$ and all the out-going 
nodes, to the right of $i$, in the gauge theory exceptional collection $\mathcal{E^Q}$ (we follow the notation of 
\cite{Herzog:2003zc} for collections) \footnote{Along this paper we follow the convention that if $C_i \cdot C_j < 0$, then 
there is an arrow in the quiver from node $i$ to node $j$. This notation differs from the one in \cite{Herzog:2003zc}
and has motivated the relabeling of the in-going and out-going sets of nodes. Both conventions are equivalent and are simply
related by charge conjugation of all the fields.}. That is

\beq
\begin{array}{cl}
C_i \cdot C_j > 0 & \Rightarrow j \mbox{ to the left of } i \\
C_i \cdot C_j < 0 & \Rightarrow j \mbox{ to the right of } i   
\end{array}
\label{well_split}
\eeq

This condition has been recently revisited in \cite{Herzog:2004qw}, where a simple geometric picture
was developed for it in terms of quivers. It relies on the identification of a special polygon in the quivers.
Finding such a polygon is not trivial, and thus its existence for every well split quiver remains a conjecture.
Here we give another, even simpler, geometric interpretation
based on $(p,q)$ webs, showing that \eref{well_split} is indeed equivalent to the absence of crossing legs.

Let us recall that the number of bifundamental fields between two nodes is equal to the intersection between the 
corresponding 3-cycles in the mirror manifold and is given by the determinant in \eref{intersection}.
Let us denote the non-trivial plane of the $(p,q)$ web as the $x-y$ plane. Then, we can think of the charges of external 
legs as defining vectors in the $x-y$ plane. Equation \eref{intersection} can be interpreted as the cross product between 
two charge vectors $\vec{C}_i$ and $\vec{C}_j$, which points in the $z$ direction. The sign of the intersection 
(equal to the sign of the angle between the two vectors, $\theta_{ij}$) just indicates whether the vector product points 
in the positive or negative $z$ direction. In this language, well split quivers correspond to $(p,q)$ webs such that, 
given any leg $i$, the rest of the legs can be separated into two sets In and Out, using the language of \cite{Feng:2002kk}, 
one to the right and one to the left of $i$ as we go around the web clockwise. These two sets are such that 
$\sin \theta_{ij}>0$ for every $j$ in In and $\sin \theta_{ij}<0$ for every $j$ in Out. This condition on angles between 
legs is clearly equivalent to the absence of crossing.

Let us pause for a moment to connect these ideas to other notions that have been developed for these quivers in the 
literature. In \cite{Herzog:2004qw}, the concept of $Ext^{1,2}$ was introduced. It turns out to be equivalent to the 
{\bf strong helix} condition and also to the absence of $Ext^3$ tachyon generating maps. Furthermore, it was also shown that $Ext^{1,2}$ 
implies well split. Putting these ideas together with the discussion above, we have

\beq
\mbox{crossing legs} \Leftrightarrow \mbox{ill split} \Rightarrow \mbox{tachyonic } Ext^3\mbox{'s} 
\eeq

We want to point out that the main virtue of interpreting the ill split quivers as arising from webs with crossing legs
comes from the fact that the web diagram indicates how the quiver has to be modified in order to correspond to a stable 
configuration given some asymptotic 7-branes.

%=====================================================
\section{A new perspective on old puzzles}
%=====================================================

\label{section_old_puzzles}

We summarize in this section a list of problematic quivers that have appeared
in the literature in the past. Some of these gauge theories were discovered when looking
at the effect of general Picard-Lefschetz monodromies on well behaved quivers. 
Others were originally constructed by a blowing-up procedure, trying to find
the quivers on the world-volume of D3-branes probing non-toric del Pezzos. These problems
appeared originally to be unrelated. Our current understanding indicates that the underlying 
common feature is that these quivers are tachyonic. We show that the presence of tachyons
can be identified in all cases using the crossing leg criterion 

%-------------------------------------------------------------
\subsection{Picard-Lefschetz monodromy versus Seiberg duality}
%-------------------------------------------------------------

Toric duality was discovered in \cite{Feng:2000mi,Feng:2001xr} as an ambiguity in the determination 
of the gauge theory on the world-volume of D3-branes probing a toric singularity. Later, it was realized that 
toric duals correspond to non-trivial realizations of Seiberg duality \cite{Feng:2001bn,Beasley:2001zp}. 
Shortly after, the question of whether the group of Seiberg duality transformations on the nodes of the 
quiver can be extended to the larger one of Picard-Lefschetz monodromies (equivalently, general mutations
in the language of exceptional collections) was raised \cite{Feng:2001bn,Feng:2002kk}, and 
whether all the gauge theories obtained this way flow to the same universality class in the IR. 

The first example of such a theory was found for the Zeroth Hirzebruch surface $F_0$ in \cite{Feng:2002kk}. The 
starting point was the theory given by the following set of $(p,q)$ 7-branes

\beq
\begin{array}{c|cccccccc}
Brane     & \ \ \ & A & \ \ \ & B & \ \ \ & C & \ \ \ & D  \\
\hline 
\hline
N_i       &       & 1  &      & 1 &       & 1 &       & 1  \\ 
\left[ p_i,q_i \right] & & [1,1] & & [1,1] & & [-3,-1] & & [1,-1]
\end{array}
\label{charges_toric_F0}
\eeq

Performing a Picard-Lefschetz monodromy of $B$ around $C$, we arrive at the new 
set of charges

\beq
\begin{array}{c|cccccccc}
Brane     & \ \ \ & A & \ \ \ & C & \ \ \ & B & \ \ \ & D  \\
\hline 
\hline
N_i       &       & 1  &      & 1 &       & 1 &       & 1  \\ 
\left[ p_i,q_i \right] & & [1,1] & & [3,1] & & [-5,-1] & & [1,-1]
\end{array}
\label{charges_fractional_F0}
\eeq

\begin{figure}[ht]
  \epsfxsize = 15cm
  \centerline{\epsfbox{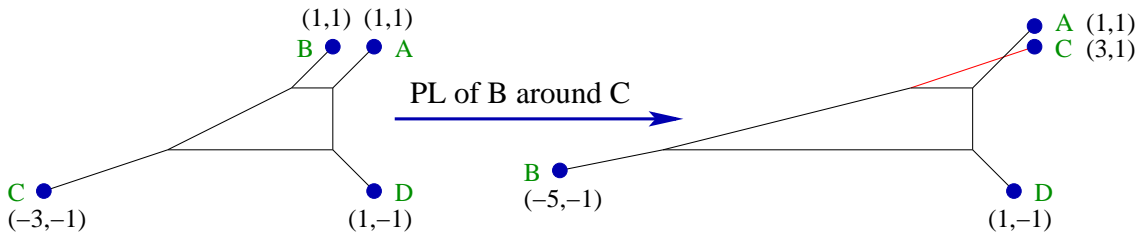}}
  \caption{$(p,q)$ webs for a normal and a fractional dual theories for $F_0$.}
  \label{webs_fractional_F0}
\end{figure}

The resulting webs are presented in \fref{webs_fractional_F0}. 

\begin{figure}[ht]
  \epsfxsize = 9cm
  \centerline{\epsfbox{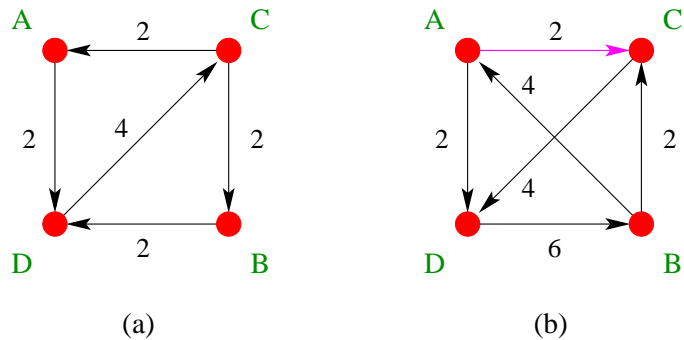}}
  \caption{Quiver diagrams for a normal and a fractional dual theories for $F_0$. The tachyonic fields arising from the           crossing legs in the web are indicated in magenta.}
  \label{quivers_F0}
\end{figure}

\fref{quivers_F0} shows the associated quivers. Although the quiver for this new model could be determined from the charges in \eref{charges_fractional_F0} using the results of \cite{Hanany:2001py}, the gauge theory was not fully specified, since the corresponding superpotential could not be determined. It is easy to verify that, if we require the theory to be conformal at the IR, not all closed loops in the quiver can appear together in the superpotential. Some of these terms must be absent in order to satisfy the vanishing of the beta function equations for their corresponding gauge and Yukawa couplings. The permitted loops can be identified by the requirement of conformal invariance. This by itself is not unusual and has been observed in other quiver theories. The new feature which is puzzling is that, in this case, it implies that some of the bifundamental fields cannot appear at all in any superpotential term. Furthermore, it was shown in \cite{Herzog:2003zc} using exceptional collections that the R-charge of gauge invariants is given by $2(n+1)$, where $n$ is the number of permutations of nodes with respect to the order they appear in the exceptional 
collection.

Using this information, even without knowing the precise form of the superpotential, one can conclude that, in some
cases, conformal invariance requires some of the R-charges to be negative.
(this is not the case for the $F_0$ example in this
section). Let us be more rigorous. The computation of R-charges using exceptional collections 
works only in those cases in which
the flavor symmetry for multiple arrows is maximal, i.e. those examples in which 
the global symmetry is such that all fields connecting a given pair of nodes have the same 
R-charge. Superpotential interactions can break these flavor symmetries. Along the rest of the 
paper, we will work under the assumption that unknown superpotentials are such that fields 
in multiple arrows have equal R-charges. The fact that negative R-charges seem to be, in some 
cases, indicative of tachyonic quivers is very suggestive, but we have to keep in mind that 
this negative values might disappear once the actual global symmetries of the theory are 
taken into account.
We will explore the possible appearance of negative R-charges in these models in Section \ref{section_negative_R} using 
the exceptional collections approach. It is interesting to note that, even in the presence of negative 
R-charges, the value of $a$ (one of the central charges of the superconformal algebra of the superconformal
field theory) remains invariant. It is conjectured that $a$ measures the number of degrees of freedom of the theory and 
obeys an analogue of the $c$-theorem in four dimensions.

%-------------------------------------------------------------
\subsection{Del Pezzo 7 and 8}
%-------------------------------------------------------------

Local mirror symmetry was used in \cite{Hanany:2001py} to obtain quivers for all del Pezzo surfaces $dP_n$. All the quivers 
were constructed using $(p,q)$ webs, starting from $dP_0$ and blowing up an increasing number of points. Known quivers were recovered 
for the toric del Pezzos $n=0$ to $3$. In addition, new quivers were proposed for the first time for $n=4$ to $8$. 
The $(p,q)$ webs constructed for $n=6$, $7$ and $8$ had crossing external legs. In \cite{Wijnholt:2002qz}, it was verified that the quivers derived 
in \cite{Hanany:2001py} for $dP_4$ and $dP_5$ were correct, after obtaining their superpotentials using exceptional 
collections. Furthermore, \cite{Wijnholt:2002qz} used an alternative 3-block quiver for $dP_6$ (whose $(p,q)$ web description does not have
crossing legs) and found its superpotential. A straightforward argument that shows that the interpretation of the $dP_7$ 
and $dP_8$ quivers of \cite{Hanany:2001py} is problematic is based on the values of R-charges. Computation of dibaryon $R$-charges determines that the minimum possible $R$-charge for 
bifundamental fields is one for $dP_7$ and two for $dP_8$. Thus, the only possible superpotential for $dP_7$ would consists of 
quadratic mass terms, while it would be impossible to construct one for $dP_8$ \cite{Herzog:2003wt}. Another fact 
is that none of these quivers can be connected by a sequence of Seiberg dualities with a healthy 3-block quiver, although they can be transformed into
one by a sequence of PL monodromies.

\begin{figure}[ht]
  \epsfxsize = 15cm
  \centerline{\epsfbox{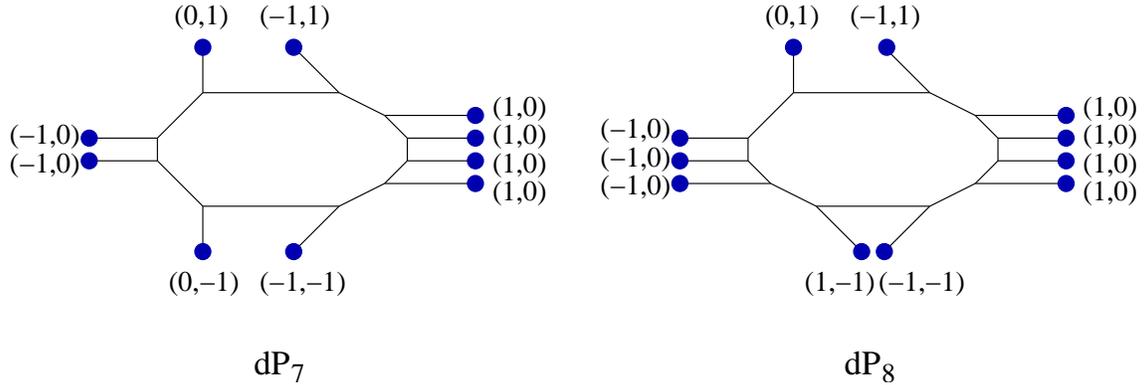}}
  \caption{$(p,q)$ webs constructed \cite{Hanany:2001py} for $dP_7$ and $dP_8$. They have crossing legs and are therefore ill split quivers.}
  \label{webs_dP7_dP8}
\end{figure}

We can now associate these difficulties, as was already noted in \cite{Franco:2004rt}, to the fact that the corresponding webs have crossing external legs as shown in \fref{webs_dP7_dP8}. 

%-------------------------------------------------------------
\subsection{Four-block models for del Pezzos}
%-------------------------------------------------------------

\label{section_4-blocks}

Our last examples correspond to a nice set of 4-block quivers, from which the $dP_1$, $dP_4$ and $dP_8$ examples were 
studied in \cite{Aspinwall:2004vm}. They can be constructed very easily using our $(p,q)$ 7-brane techniques. The 
7-brane backgrounds that correspond to the mirror of local $dP_n$ surfaces are given by \cite{Hanany:2001py}
                                                                   
\beq
\begin{array}{c|cccccccccccc}
          & \ \ \ & A_1 & \ \ \ & \ldots & \ \ \ & A_n & \ \ \ &  B & \ \ \ & C & \ \ \ &  D \\
\hline 
\hline
\left[ p_i,q_i \right] & & [1,0] & & \ldots & & [1,0] & & [2,-1] & & [-1,2] & & [-1,-1]
\end{array}
\eeq
                                                                                
This configuration does not satisfy $\sum_i(p_i,q_i)=0$ and thus does not
correspond to an anomaly free quiver with all ranks equal to $N$. From this configuration, we can 
immediately construct another one that corresponds to a 4-block quiver with all gauge groups equal, by 
using PL monodromies to move the B brane all the way to the left of the $n$ $A_i$ branes. The $(p,q)$ charge 
of $A_i$ is $(1,0)$ and its monodromy can be computed from \eref{monodromy} to be
                                                                             
\beq
\begin{array}{ccc}
M(1,0)=\left(\begin{array}{rcr} 1 & \ & 1 \\ 0 & &  1 \end{array}\right) &
\ \ \ \ \ &
M(1,0)^n=\left(\begin{array}{rcr} 1 & \ & n \\ 0 & &  1 \end{array}\right)
\end{array}
\eeq
 
Then, the final collection of 7-branes becomes

\beq
\begin{array}{c|cccccccccccc}
          & \ \ \ & B & \ \ \ & A_1 & \ \ \ & \ldots & \ \ \ & A_n & \ \ \ & C & \ \ \ & D \\
\hline 
\hline
N_i       &       & 1 &       &   1 &       & \ldots &       &   1 &       & 1 &       & 1 \\ 
\left[ p_i,q_i \right] & & [2-n,-1] & & [1,0] & & \ldots & & [1,0] & & [-1,2] & & [-1,-1]
\end{array}
\label{charges_4-block}
\eeq

\begin{figure}[ht]
  \epsfxsize = 14cm
  \centerline{\epsfbox{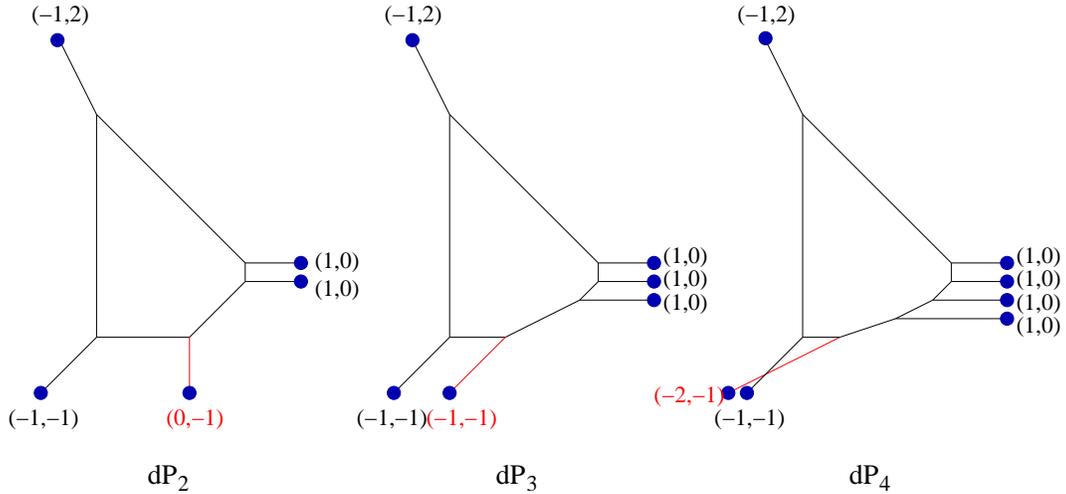}}
  \caption{$(p,q)$ webs for some 4-block models for $dP_2$ to $dP_4$. The $dP_3$ model in this family is actually a 3-block
           theory. Crossing legs appear for $dP_n$, with $n>3$.}
  \label{webs_4-blocks}
\end{figure}

In order to illustrate how the crossing legs appear in these examples, we present in \fref{webs_4-blocks} the $(p,q)$ webs 
for $dP_2$ to $dP_4$.

\begin{figure}[ht]
  \epsfxsize = 14cm
  \centerline{\epsfbox{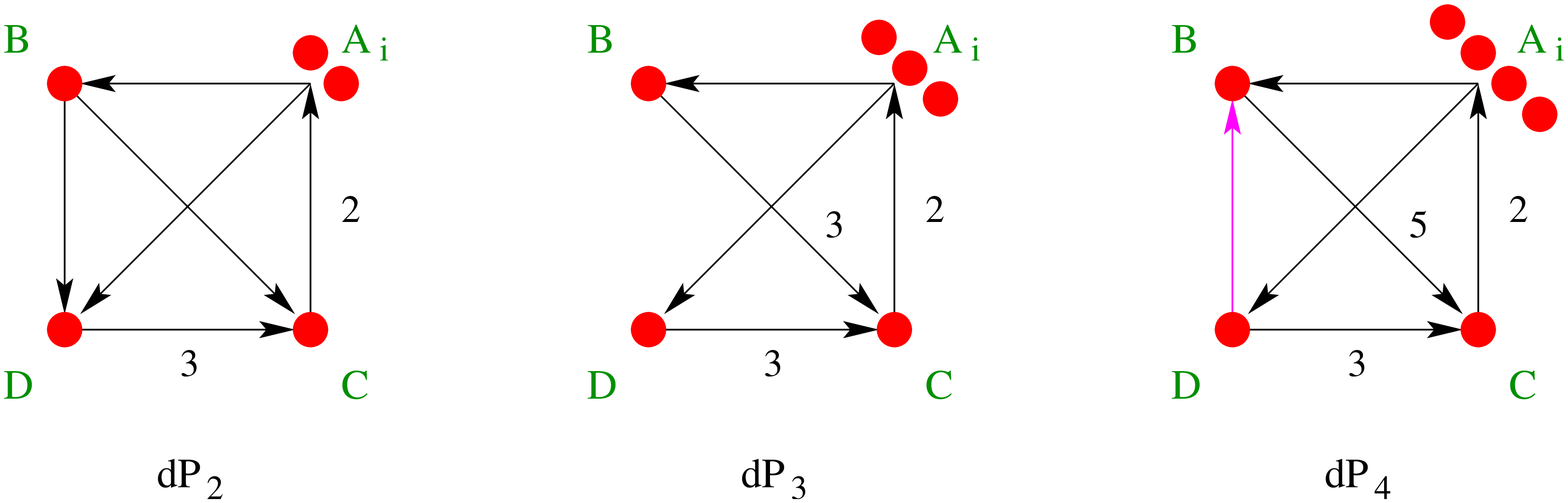}}
  \caption{Quiver diagrams for the 4-block models whose webs are shown in \fref{webs_4-blocks}. The tachyonic fields arising from the crossing legs in the web are indicated in magenta.}
  \label{quivers_4-blocks}
\end{figure}

Labeling rows and columns in the order $(B,A_i, C, D)$, we get the following intersection matrix
                                                                                                                                                           
\beq
\mathcal{I}=\left(\begin{array}{rcrcrcr} 
 0    & \ \ &  1 & \ \ & 3-2n & \ \ & n-3 \\
-1    & \ \ &  0 & \ \ &    2 & \ \ &  -1 \\
 2n-3 & \ \ & -2 & \ \ &    0 & \ \ &   3 \\
 3-n  & \ \ &  1 & \ \ &   -3 & \ \ &   0
\end{array}\right)
\eeq
which can be encoded in the quiver diagrams in \fref{quivers_4-blocks}.

From the $(p,q)$ charges in \eref{charges_4-block}, we see that the the crossing legs appear for $n>3$. Furthermore, 
the intersection number between the crossing legs is

\beq
B \cdot D=n-3
\eeq
in precise agreement with results from the computation of the dimensions of $Ext^3$'s of \cite{Aspinwall:2004vm}
\footnote{The subsection in which this result appeared was removed in the second version of \cite{Aspinwall:2004vm}, but the computation remains valid.}.

%===========================================================
\section{Getting rid of the tachyons: going to higher genus}
%===========================================================

\label{section_higher_genus}

We have discussed in Section \ref{section_well_split} how tachyonic fields can be identified by looking for crossing 
legs in the associated $(p,q)$ webs. We illustrated this statement in Section \ref{section_old_puzzles}, in a large
set of examples. According to the proposal we outlined in Section \ref{section_crossing_legs}, whenever we arrive at a 
configuration with crossing external legs by a PL monodromy transformation, the new set of stable branes corresponds to a higher 
genus resolution of the geometry. We now implement our proposal in an explicit example, computing the final quiver after 
the monodromy.

Let us start from the following set of $(p,q)$ branes 

\beq
\begin{array}{c|cccccccc}
Brane     & \ \ \ & 1 & \ \ \ & 2 & \ \ \ & 3 & \ \ \ & 4  \\
\hline 
\hline 
N_i       &       & 1  &      & 1 &       & 1 &       & 1  \\ 
\left[ p_i,q_i \right] & & [0,-1] & & [1,-2] & & [1,0] & & [-2,3]
\end{array}
\label{charges_toric_dP1}
\eeq

\begin{figure}[ht]
  \epsfxsize = 9cm
  \centerline{\epsfbox{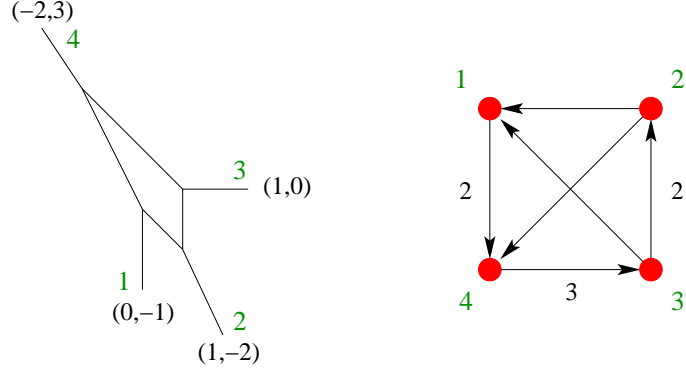}}
  \caption{$(p,q)$ web and quiver diagram for one of the phases of $dP_1$.}
  \label{web_quiver_dP1}
\end{figure}

This configuration corresponds to $dP_1$. We show the $(p,q)$ web and the associated quiver in \fref{web_quiver_dP1}.

Let us now perform a PL monodromy of node $2$ to the right around $3$. This transformation does not correspond 
to a Seiberg duality. The resulting configuration can be computed using \eref{monodromy} and is given by

\beq
\begin{array}{c|cccccccc}
Brane     & \ \ \ & 1 & \ \ \ & 3 & \ \ \ & 2 & \ \ \ & 4  \\
\hline 
\hline 
N_i       &       & 1  &      & 1 &       & 1 &       & 1  \\ 
\left[ p_i,q_i \right] & & [0,-1] & & [-1,0] & & [3,-2] & & [-2,3]
\end{array}
\label{charges_monodromy_dP1}
\eeq

These charges give rise to the following intersection matrix

\beq
\mathcal{I}=\left(\begin{array}{rrrr}
 0 &  3 & -1 & -2 \\
-3 &  0 & -2 &  5 \\  
 1 &  2 &  0 & -3 \\
 2 & -5 &  3 &  0 
\end{array}\right).
\label{intersection_mutation_dP1}
\eeq

\begin{figure}[ht]
  \epsfxsize = 9cm
  \centerline{\epsfbox{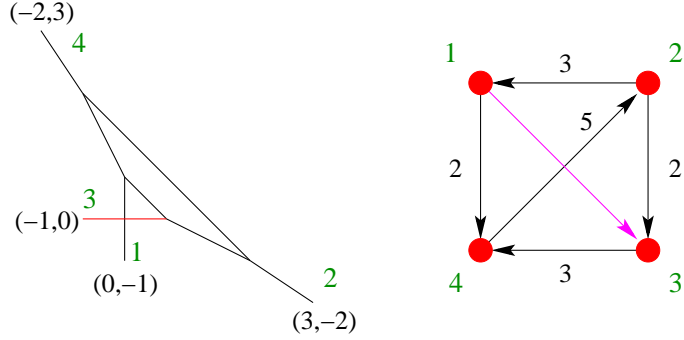}}
  \caption{$(p,q)$ web and quiver diagram obtained by PL monodromy of node 2 around 3 on the configuration in \fref{web_quiver_dP1}. The crossing of legs 1 and 3 give rise to tachyonic bifundamental fields between the corresponding nodes that we indicate in magenta.}
  \label{web_quiver_monodromy_dP1}
\end{figure}

The resulting web and quiver are shown in \fref{web_quiver_monodromy_dP1}. We have indicated in magenta the tachyonic 
bifundamental fields from the intersection of the crossing legs.

This monodromy transformation can be also seen as a mutation in the corresponding exceptional collection 
$\mathcal{E^Q}=(E_1,E_2,E_3,E_4)$. The Chern characters of the initial sheaves 
$ch(E_i)=(r(E_i)+c_1(E_i)+ch_2(E_i))$ are

\beq
\begin{array}{rc}
ch(E_1): & (-1,H-E_1,0) \\ 
ch(E_2): & (0,E_1,-1/2) \\
ch(E_3): & (2,-H,-1/2)  \\
ch(E_4): & (-1,0,0)
\end{array}
\label{ex_collection_toric_dP1}
\eeq

Performing a left right mutation of $E_2$ around $E_3$, $E_2 \rightarrow R_{E_3}E_2$, we get

\beq
\begin{array}{rc}
ch(E_1): & (-1,H-E_1,0) \\ 
ch(E_3): & (-2,H,1/2)  \\
ch(E_2): & (4,-2H+E_1,-3/2) \\
ch(E_4): & (-1,0,0)
\end{array}
\label{ex_collection_mutation_dP1}
\eeq
where we have inverted the sign of $ch(E_2)$. The intersection matrix computed with these charges
is in agreement with \eref{intersection_mutation_dP1}, computed using the $(p,q)$ charges of equation 
\eref{intersection_mutation_dP1}.

\begin{figure}[ht]
  \epsfxsize = 12cm
  \centerline{\epsfbox{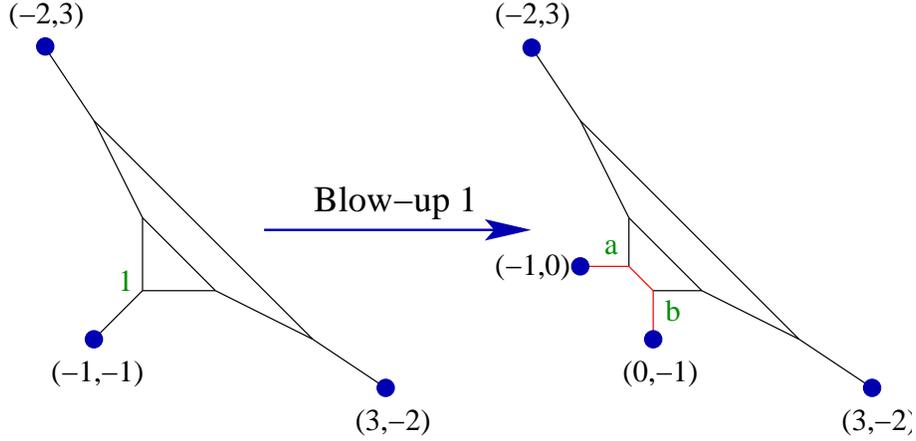}}
  \caption{Connection by a blow-up between the $(p,q)$ web for the $\IC^3/\IZ_5$ orbifold and the one for a fractional
           Seiberg dual of $dP_1$ corresponding to the resolution of the crossing legs.}
  \label{webs_blow_up}
\end{figure}

The tachyonic quiver of \fref{web_quiver_monodromy_dP1} has to be replaced with a new, stable one. 
The stable quiver corresponds to a genus 2 $(p,q)$ web that results after the external legs cross.
The number of gauge groups is given by the number of nodes in the web. Therefore, we have six gauge groups
in this case. The $(p,q)$ web indicates that this theory corresponds to a blow-up of $\IC^3/\IZ_5$ as 
shown in \fref{webs_blow_up} \footnote{The $(p,q)$ web diagram for $\IC^3/\IZ_5$ can be found, for example, in \cite{Feng:2002kk}.}. 

\begin{figure}[ht]
  \epsfxsize = 5cm
  \centerline{\epsfbox{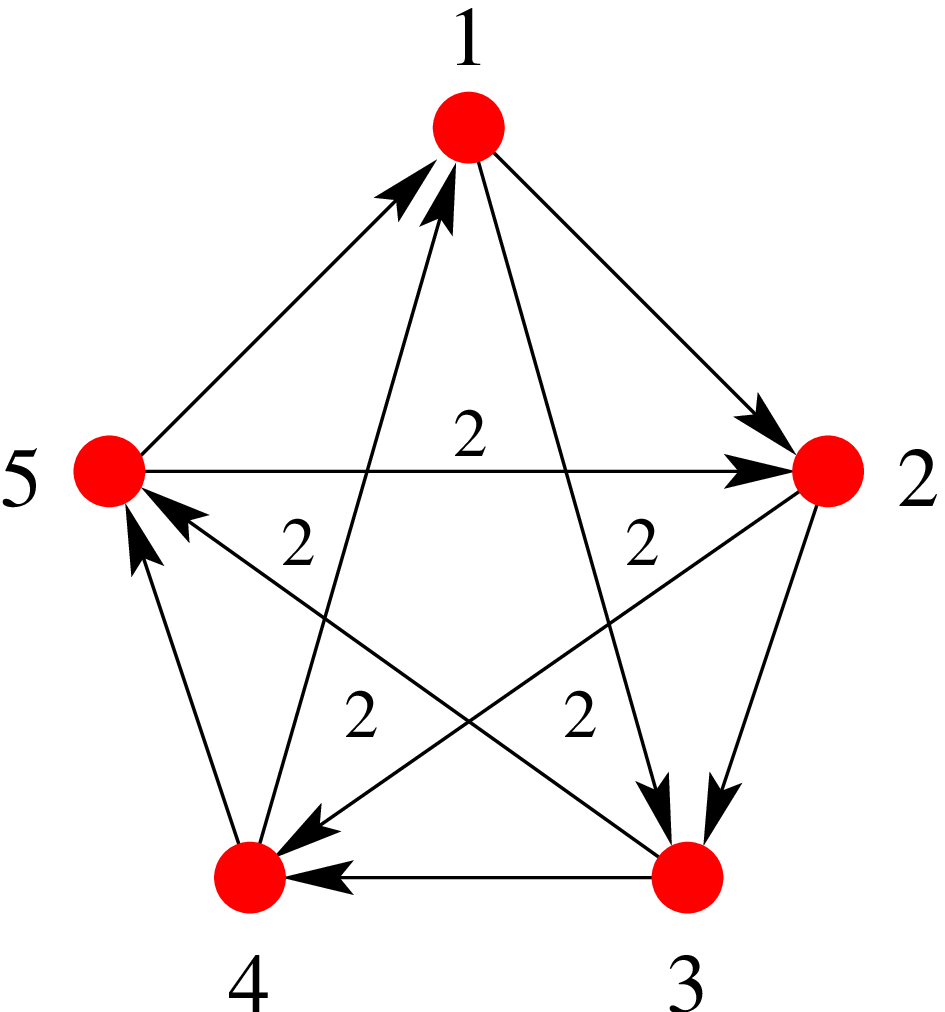}}
  \caption{Quiver diagram for the $\IC^3/\IZ_5$ orbifold.}
  \label{quiver_Z5}
\end{figure}

The gauge theory on D-branes probing an orbifold can be determined using the standard techniques of \cite{Douglas:1996sw}.
For D3-branes over the $\IC^3/\IZ_5$ orbifold, we have the quiver presented in \fref{quiver_Z5} 
with superpotential

\beq
W=\sum_{i=1}^5 (X_{i,i+2} Y_{i+2,i+4} Z_{i+4,i}-Y_{i,i+2} X_{i+2,i+4} Z_{i+4,i})
\label{W_Z5}
\eeq
where the nodes are identified modulo 5.

\begin{figure}[ht]
  \epsfxsize = 11cm
  \centerline{\epsfbox{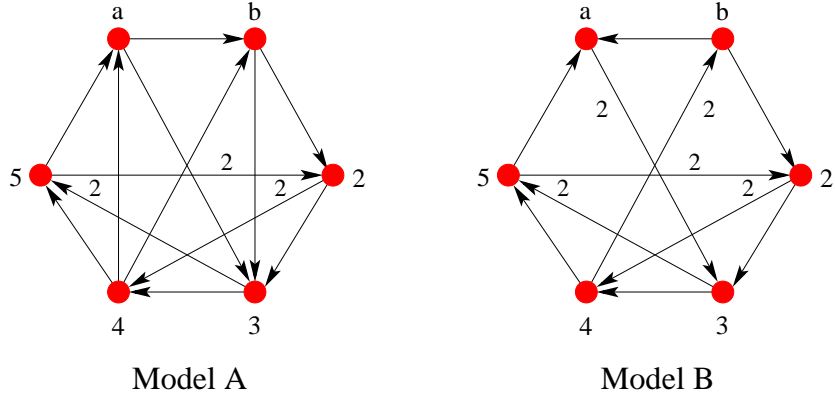}}
  \caption{Quiver diagrams for the two inequivalent toric blow-ups of $\IC^3/\IZ_5$.}
  \label{quivers_g2}
\end{figure}

\begin{figure}[ht]
  \epsfxsize = 9cm
  \centerline{\epsfbox{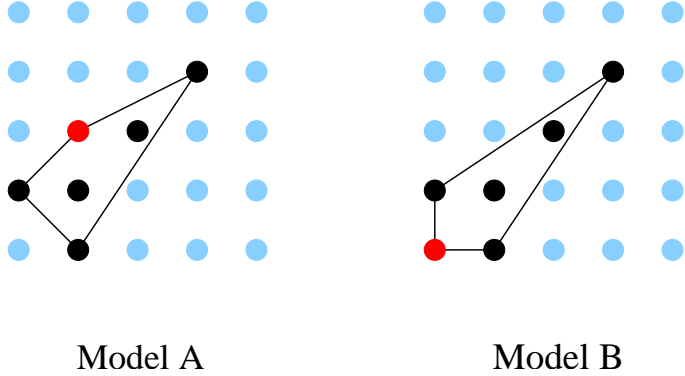}}
  \caption{Toric diagrams for the two inequivalent blow-ups of $\IC^3/\IZ_5$. We have indicated the new node in the toric diagram in red.}
  \label{toric_diagrams}
\end{figure}

A geometric blow-up corresponds in the gauge theory on the brane probes to an unhiggsing \cite{Feng:2002fv}. 
Using \fref{webs_blow_up}, we can identify the correspondence between the nodes of the $\IC^3/\IZ_5$ and those
of the new geometry. Node $1$ of the orbifold comes from the higgsing of the nodes $a$ and $b$. 
Furthermore, the blow-up condition implies that there is exactly one bifundamental field between 
these two nodes. Since the underlying geometry is an affine toric variety, every bifundamental field 
must appear exactly twice in the superpotential \cite{Feng:2000mi}. These observations
determine that there are only two possible unhiggsed theories. This is in accordance
with the fact that there are only two possible inequivalent toric blow-ups of the $\IC^3/\IZ_5$ geometry. These blow-ups are
shown in \fref{toric_diagrams}. The two quivers obtained in this way are presented in \fref{quivers_g2}.

% This quiver corresponds to the following intersection matrix

% \beq
% \mathcal{I}=\left(\begin{array}{rrrrrr} 
% 0 & -1 & 0 & -1 & 1 & 1 \\
% 1 & 0 & -1 & -1 & 1 & 0 \\
% 0 & 1 & 0 & -1 & -2 & 2 \\
% 1 & 1 & 1 & 0 & -1 & -2 \\
% -1 & -1 & 2 & 1 & 0 & -1 \\
%  -1 & 0 & -2 & 2 & 1 & 0
% \end{array}\right)
% \eeq
% where rows and columns are arranged in the following order: $(a,b,2,3,4,5)$. The superpotential for this
% model is

The superpotential for Model A is

\beq
\begin{array}{rl}
W_A=& X_{52}Z_{23}Y_{35} - Y_{52}Z_{23}X_{35} + X_{24} Z_{45} Y_{52} - Y_{24} Z_{45} X_{52}+X_{35}X_{5a}X_{a3}-X_{a3}Z_{34}X_{4a} \\
 +& X_{b3} Z_{34} X_{4b}-X_{4b}X_{b2}X_{24}+X_{4a}X_{ab}X_{b2}Y_{24}-Y_{35}X_{5a}X_{ab}X_{b3}
\end{array}
\label{W_g2A}
\eeq
while the one for Model B is 

\beq
\begin{array}{rl}
W_B& = X_{35} Z_{5a} Y_{a3}-Y_{35} Z_{5a} X_{a3}+X_{24} Z_{45} Y_{52}-Y_{24} Z_{45} X_{52} \\
  & + X_{52} Z_{23} Y_{35}-Y_{52} Z_{23} X_{35}+X_{4b} Z_{b2} Y_{24}-Y_{4b}Z_{b2} X_{24} \\
  & + X_{a3} Z_{34} Y_{4b} X_{ba}-Y_{a3} Z_{34} X_{4b} X_{ba}
\end{array}
\label{W_g2B}
\eeq

We observe that Model B has an $SU(2)$ global symmetry, which is broken in Model A by the superpotential.

It is straightforward to check that both quivers in \fref{quivers_g2}, with their associated superpotentials given by \eref{W_g2A} and 
\eref{W_g2B}, reduce to the ones of $\IC^3/\IZ_5$ by higgsing. In order to verify which of the gauge theories is the one that we are looking
for, which corresponds to the second $(p,q)$ web in \fref{webs_blow_up}, we compute their moduli spaces. The corresponding toric diagrams 
are shown in \fref{toric_diagrams}, from which we conclude that Model B is the theory we are pursuing.

%=====================================================
\section{Negative R-charges}
%=====================================================

\label{section_negative_R}

We have discussed the existence of tachyons in fractional Seiberg dual quivers and its connection
to the crossing legs in the associated $(p,q)$ web. Let us explore now another characteristic
signature of some of these theories.

A peculiarity of some of the fractional dual models is the existence of negative R-charges. This fact has already
been noticed in the past \cite{Herzog:2003zc}. Let us consider one of such examples. A practical way to compute R-charges of del
Pezzo quivers, based solely on information in the quiver, was derived in \cite{Herzog:2003dj} from exceptional collections. It is 
given by

\beq
R(X_{ij})={2 \over (9-n) d_i d_j} \times \left\{\begin{array}{ll} 
S_{ij}^{-1}   & \mbox{if } S_{ij}^{-1}\neq 0 \\
2-S_{ij}^{-1} & \mbox{if } S_{ij}^{-1}= 0
\end{array}\right.
\label{R-charges}
\eeq
where $S$ is the incidence matrix of the quiver. $S$ is upper triangular and such that $S_{ij}=\mathcal{I}_{ij}$ for 
$i<j$ and $S_{ii}=1$. This method is closely related to the one based on the calculation of R-charges for dibaryon
operators of \cite{Intriligator:2003wr}. An alternative, more general, procedure is to require the beta functions for gauge 
and superpotential couplings to vanish and then to fix the remaining freedom using the principle of $a$-maximization 
\cite{Intriligator:2003jj}.

Let us apply \eref{R-charges} to the 4-block del Pezzo models discussed in Section \ref{section_4-blocks}. The incidence 
matrix for these models is

\beq
S=\left(\begin{array}{rcrcrcr} 
1 & \ \ & -1 & \ \ & 2n-3 & \ \ & 3-n \\
0 & \ \ &  1 & \ \ &   -2 & \ \ &   1 \\
0 & \ \ &  0 & \ \ &    1 & \ \ &  -3 \\
0 & \ \ &  0 & \ \ &    0 & \ \ &   1
\end{array}\right)
\eeq
from which we obtain the following set of R-charges

\beq
\begin{array}{lcl}
n\leq 1: \ R_{BC}=5/4-1/2 n>0 & \ \ \ \ & n>1: \ R_{CB}=3/4+1/2 n \\
n\leq 3: \ R_{BD}=5/4n-3/4>0  &         & n>3: \ R_{DB}=11/4-5/4 n<0 \\
R_{CA}=1/2                    &         & R_{AD}=3/4 \\
R_{DC}=3/4                    &         & 
\end{array}
\eeq

We see that $R_{DB}$ becomes negative for $n>3$, exactly when the D and B legs start crossing. In this case, gauge 
invariant dibaryon states with negative R-charge, and hence negative conformal dimension, can be constructed 
by antisymmetrizing $N$ copies of the negative R-charge $X^i_{DB}$ bifundamental fields. This fact would
violate unitarity.

%=====================================================
\section{Conclusions}
%=====================================================

The precise significance of general Picard-Lefschetz monodromy transformations on 3-cycles that do not give rise
to Seiberg dualities in the field theories on D3-branes probing the mirror manifold has been obscure in the past.
The same problem has appeared as an unclear interpretation of general mutations in the exceptional collection approach
to quivers for D-branes on singularities. 

Based on the $(p,q)$ web description of singularities we have developed a novel perspective into 
the tachyonic quivers that arise after fractional Seiberg duality. In Section \ref{section_well_split}, we proved the equivalence
between the ill split condition of exceptional collections and the crossing of external legs in the associated
$(p,q)$ web. This new interpretation makes the identification of tachyonic ill split quivers straightforward. We
re-examined various previously problematic quivers in Section \ref{section_old_puzzles} under this new light.
Another advantage of this approach is that it indicates the set of stable marginally bound states of D-branes after monodromies are performed.
This information is given by specifying the corresponding quiver and superpotential of a higher genus $(p,q)$ web.
For ill split quivers, the corresponding web gets an increase of the genus, i.e. an increase in the number of collapsing 4-cycles in the singularity.

In Section \ref{section_higher_genus}, we applied our proposal to a specific example obtained after performing a Picard-Lefschetz monodromy
on a $dP_1$ quiver. The new model corresponds to a genus 2 $(p,q)$ web. Using the $(p,q)$ web to relate the new theory to the $\IC^3/\IZ_5$ orbifold by a blow-up, we managed to determine both the quiver and its superpotential after the monodromy. As a bonus, this theory represents the first example in which the gauge theory on a stack of D3-brane probes has been determined for a non-orbifold toric singularity with more than one collapsing 4-cycles. We are hopeful that the techniques discussed in this paper will permit a systematic study of quivers for singularities of genus greater than one. 

A clear challenge is how to understand the genus increase and how to identify the stable quiver after monodromy within the frameworks of exceptional collection and of the derived category of quiver representations.

We hope to perform a detailed application of the ideas presented in this 
note in the near future.

%===========================================================
\section{Acknowledgements}
%===========================================================

We would like to thank P. Aspinwall, C. Herzog, P. Kazakopoulos and A. Sen for illuminating discussions. We are specially 
grateful to Y. H. He. for useful conversations and collaboration on a related project. S. F. would like to thank
the organizers of the Cargese Summer School ``String Theory: From Gauge Interactions to Cosmology" for their hospitality 
while this work was being completed. This Research was supported in part by the CTP and LNS of MIT and the U.S. Department 
of Energy under cooperative research agreement \# DE-FC02-94ER40818, and by BSF -- American-Israeli
Bi-National Science Foundation. A. H. is also indebted to the Reed Fund Award and 
a DOE OJI Award.

%%%%%%%%%%%%%%%%%%%%%%%%%%%%%%%%%%%%%%%%%%%%%%%%%%%%%%%%%%%%%%%%%%%%%%%%%%%%%%%%%%%%%%%%%%%%%%%%%%%%%%%%%%%%%%%%%%%%%%%%%%%%%%%%%%%%%%%%%%%%%%%%%%%%%%%

\bibliographystyle{JHEP}

\end{document}